\documentclass[natbib]{emulateapj}
\usepackage[]{natbib}

\usepackage[]{verbatim}

\newcommand{\kms}{km\ s$^{-1}$}

\newcommand{\vsini}{$v$\,sin\,$i$}

\shorttitle{Sco Cen}
\shortauthors{Bubar et al.}

\begin{document}

\title{Lithium in the Upper-Centaurus Lupus and Lower-Centaurus Crux Subgroups of Scorpius-Centaurus}
\author{Eric J. Bubar\altaffilmark{1}}
\affil{Department of Biology and Physical Sciences, Marymount University, 2807 N. Glebe Road,
Arlington, VA 22207}
\altaffiltext{1}{Department of Physics and Astronomy, University of Rochester,
    P.O. Box 270171, Rochester, NY 14627-0171}
\author{Marc Schaeuble, Jeremy R. King }
\affil{Department of Physics and Astronomy, Clemson University,
    Clemson, SC 29630-0978}
\author{Eric E. Mamajek}
\affil{Department of Physics and Astronomy, University of Rochester,
    P.O. Box 270171, Rochester, NY 14627-0171}
\author{John R. Stauffer}
\affil{Spitzer Science Center, Caltech, Pasadena, CA 91125}
\begin{abstract}

We utilize spectroscopically derived model atmosphere parameters and the \ion{Li}{1}
$\lambda6104$ subordinate line and the $\lambda6708$ doublet to derive lithium
abundances for 12 members of the Upper-Centaurus Lupus (UCL)
and Lower-Centaurus Crux (LCC) subgroups of the Scorpius Centaurus OB
Association.  The results indicate any intrinsic Li scatter
in our 0.9-1.4 $M_{\odot}$ stars is limited to ${\sim}0.15$ dex, consistent
with the lack of dispersion in ${\ge}1.0$ $M_{\odot}$ stars in the 100 Myr Pleiades 
and 30-50 Myr IC 2391 and 2602 clusters.  Both ab
initio uncertainty estimates and the derived abundances themselves indicate that the
$\lambda$6104 line yields abundances with equivalent or less scatter than is found from the
$\lambda$6708 doublet as a result of lower uncertainties for the subordinate
feature, a result of low sensitivity to broadening in the subordinate feature.  Because NLTE corrections are less susceptible to changes in surface
gravity and/or metallicity for the 6104 {\AA} line, the subordinate Li feature
is preferred for deriving lithium abundances in young Li-rich stellar
association stars with $T_{\rm eff}{\ge}5200$ K.  At these temperatures, we find no
difference between the Li abundances derived from the two \ion{Li}{1} lines. For
cooler stars, having temperatures at which main-sequence dwarfs show abundance
patterns indicating overexcitation and overionization, the ${\lambda}6104$-based Li 
abundances are ${\sim}0.4$ dex lower than those derived from the $\lambda$6708 doublet.  
The trends of the abundances from each feature with $T_{\rm eff}$ suggest that this
difference is due to (near)UV photoionization, which in NLTE preferentially
ionizes Li atoms in the subordinate 2\emph{p} state relative to 2\emph{s} resonance
line state due to opacity effects.  Consequently, this overionization of Li in the 2\emph{p} state, 
apparently not adequately accounted for in NLTE corrections, weakens the $\lambda$6104
feature in cooler stars.  Accordingly, the $\lambda$6708-based
abundances may give more reliable estimates of the mean Li abundance in
cool young stars.  Our mean Li abundance, log $N$(Li)$=3.50{\pm}0.07$ is ${\sim}0.2$
dex larger than the meteoritic value.  While stellar models suggest that Li depletion of
at least 0.4 dex, and possibly much larger, should have occured in our
lowest mass star(s), our Li abundances show no decline with decreasing mass indicative of
such depletion.

\end{abstract}


\section{INTRODUCTION}

The identification
of the lower mass members of Scorpius-Centaurus, the nearest OB Association to 
our Sun, has provided a rich source of nearby (within 150 pc) sun-like pre-MS stars for understanding early stellar
evolution.  Sco-Cen comprises three primary subgroups: Upper-Sco (US),
Upper-Centaurus Lupus (UCL) and Lower-Centaurus Crux (LCC) with ages of $\sim$10 Myr \citep{Pecaut2011}, 
$\sim$15 Myr and $\sim$16 Myr \citep{Preibisch2008}, respectively.  One of the key pieces of the puzzle 
in understanding young stars and their internal structure lies within understanding their lithium abundances.  
Since lithium is depleted during the pre-main sequence (by 0.30-1.0 dex for a 1 M$\odot$, solar composition star \citep{PT02}),
it provides a strong constraint on internal stellar structure and has provided a valuable diagnostic of
youth, yet several puzzling questions remain regarding lithium depletion.

Historically, lithium has been a valuable target for much abundance work, particularly regarding
open clusters.  Stars with $M>1$ M$_{\odot}$ in the 100 Myr Pleiades cluster and in the 30-50 
Myr IC 2391 and 2602 clusters evince no Li star-to-star Li dispersion above the uncertainties 
(Figures 3 and 7 of King et al.~2000; Figure 4 of Randich et al.~2001).  One of the most vexing problems lies 
in the star-to-star dispersion in lithium abundances for $M{\le}0.8$ M$_{\odot}$ in  
the Pleiades cluster \citep{Soderblom1993,KSHP10}, a result at odds with standard stellar models and the lack of abundance 
scatter in lower mass IC 2391 and 2602 stars.  In addition Li dispersion of the size seen in the Pleiades in not apparent in 
lower mass stars in clusters of older ages \citep[e.g. Hyades:][]{Soderblom1995}.  

\citet{Russell1996} examined this Pleiades dispersion by exploring lithium abundances derived from
two lithium features, the strong doublet at 6708 {\AA} (2\emph{p}-2\emph{s} transition)
and the weaker subordinate feature at 6104 {\AA} (3\emph{d}-2\emph{p} triplet transition)
\citep[Grotrian diagram in Figure 3 of]{Carlsson1994}.
He found that abundances from the subordinate line were greater than those from the resonance 
doublet and that less scatter was observed in lithium abundances derived from equivalent widths of 
the 6104 {\AA} line ($\pm$0.1 dex) when compared with those derived from the 6708 {\AA} line 
($\pm$0.2 dex).  \citet{Russell1996} suggested that a larger-scale survey of lithium in the Pleiades using 
the 6104 {\AA} feature might explain the Pleiades Li dispersion as an artifact of 
spurious ${\lambda}6708$-based abundances.  

\citet{Ford2002} reexamined the work of \citet{Russell1996} using spectra of 11 late-G and early-K 
Pleiads and suggested that improper accounting of an \ion{Fe}{2} blend in the 6104 {\AA} region was 
likely responsible for both the decreased spread and enhanced abundances they reported.  By utilizing 
spectral synthesis of 
the 6104 {\AA} which accounted for this \ion{Fe}{2} blending feature, the spread in abundances derived 
by \citet{Ford2002} was equal for both the $\lambda$6104 and $\lambda$6708 lines, indicating a 
genuine abundance spread for the Pleaides.  In addition, while some abundances from the subordinate 
line agreed with those from the 6708 doublet, there were 4 stars which showed clearly larger abundances 
from $\lambda$6104, which \citet{Ford2002} suggested might be the result of spot coverage on those stars.

Most recently, \citet{KSHP10} revisited lithium in the Pleiades and compared lithium abundances derived 
from both $\lambda$6104 and $\lambda$6708.  They also found that the scatter in 6104 {\AA} abundances 
persisted and was comparable to that observed in the 6708 {\AA} results.  However, contrary to  
\citet{Ford2002}, they found little difference between abundances from the two lines; the source of 
these discrepant conclusions is unclear.  Salient questions that remain are which line gives the most 
reliable Li abundances, and whether the Pleiades Li scatter is a relic 
of differential PMS depletion already in place by the ${\sim}15$ Myr age of our stars.  While our sample's
mass range limits its utility for exploring the latter, it is well-suited for former.

Recent work suggests that ZAMS-to-Hyades age dwarfs with $T_{\rm eff}{\le}5200$ K show abundance 
anomalies which mimic the effects of overionization \citep{Sch03}.   The reality of overionization in such stars 
can be explored using the  two \ion{Li}{1} features alone by exploiting the NLTE elements of \ion{Li}{1} line 
formation at such temperatures.  Due to a conspiracy of the photoionizing energies and the UV line opacity, 
the 6104 {\AA} lower energy level (2\emph{p}) can be depopulated relative to the 6708 {\AA} ground state 
(2\emph{s}) \citep{Carlsson1994}.   
If the extent of this effect is not accounted for in NLTE calculations, the ${\lambda}6104$ feature, weakened 
relative to the resonance line, would yield lower abundances.   Might the Li-rich stars in Sco-Cen yield
6104 {\AA} lines strong enough to measure abundances of sufficient accuracy to search for such an offset?

Motivated by these questions, we have derived lithium for a sample of UCL and 
LCC members of Sco-Cen.  Being young ($\sim$ 15 Myr), the lithium abundances in our stars are
necessarily large, which decreases the uncertainty in spectral synthesis of the weaker 6104 {\AA} lines which may
affect the aforementioned studies of the Pleiades.  For our mass range, the expected modest depletion factors
allow us to determine abundances through spectral synthesis of both the $\lambda$6104 feature and the 
$\lambda$6708 doublet.  We therefore derive lithium for a sample of 12 UCL and LCC members from these 
two lithium features.  While our mass range limits the viability
of answering questions regarding depletion mechanisms, we are able to robustly constrain the lithium
abundances of our stars using the traditionally weaker $\lambda$6104 subordinate line as well as the
$\lambda$6708 doublet.  We compare the results from the two lines, analyze the differences in terms of
potential NLTE effects, make recommendations for which features to use in determining lithium abundances 
in other young clusters and associations, and determine a mean abundance of lithium for the UCL/LCC
subgroups of Sco-Cen.

\section{DATA, OBSERVATIONS AND ANALYSIS}

\subsection{Spectroscopic Observations and Reductions}

The sample explored here is a subset of the sample of \citet{Mamajek2002}, who identified
members of the UCL and LCC subgroups of the Scorpius-Centaurus OB Association using x-ray, proper-motions
and color-magnitude selection.  Optical spectroscopy of the sample was obtained on June 14-17, 
2002 with the CTIO 4 meter Blanco 
telescope and the echelle spectrograph 
with 31.6 l/mm grating  and a 2048X2048 CCD detector.  The 0.8\arcsec\, slit width yielded a resolution of 
R $\sim$ 40,000 with a typical S/N of 60-80 per summed pixel
in exposure times of $\approx$100-300 s.  The spectra 
have incomplete wavelength coverage extending from approximately 5800 {\AA}  to 7800 {\AA}, across
45 orders.  
The spectra have been reduced using standard routines in the {\sf echelle} package of 
IRAF\footnote{IRAF is distributed by the National Optical Astronomy Observatories, 
which are operated by the Association of Universities for Research in Astronomy, Inc., 
under cooperative agreement with the National Science Foundation.}.  These include 
bias correction, flat-fielding, scattered light correction, order extraction, and 
wavelength calibration.
We have selected stars which meet three criteria for this study: 1) spectra are
higher S/N ($\sim$60-80) 2) stars are apparently slower rotators (\vsini $\le$ 20
\kms) and 3) spectra showed no evidence of multiplicity in their cross-correlation peaks 
when correlated with a template solar spectrum taken 
with the same instrument.  Further details of the velocity and multiplicity results
can be found in \citet{epaper}.

\section{ABUNDANCE ANALYSIS AND UNCERTAINTIES}

Abundances from the \ion{Li}{1} resonance feature at 6708 {\AA} and the 1.8 eV subordinate 
features at 6104 {\AA} were derived from spectral synthesis using an updated version of the 
LTE analysis package {\sf MOOG} \citep{Sneden73}.  Stellar parameters were derived using excitation and
ionization balance following the approach described in \citet{Bubar2010} and are listed in Table 1. Model atmospheres
corresponding to these parameters were interpolated from the Kurucz ATLAS9 
grids\footnote{http://kurucz.harvard.edu/grids.html}.  The ${\lambda}6104$ linelist was that
described and used by \citet{KSHP10}. The ${\lambda}6708$ region linelist was that used by 
\cite{King97} updated using the CN features from \citet{Mandell04} and additional atomic data
from Kurucz\footnote{http://kurucz.harvard.edu} and VALD \citep{Pis95,Kupka99}.  
\marginpar{Tab1}

For MML 28, comparison of the synthesis with the observed data in the 6708 {\AA} region suggested
a differential radial velocity shift between Li and other metal lines that could not be 
rectified with variations in smoothing or input abundances in a fashion that yielded acceptable 
fits to the 6104 {\AA} region.  The inferred differential shift of the 6708 {\AA} feature 
was blueward by 0.7 km/s ($-0.016$ {\AA}).  \citet{Cayrel07} found that 3D NLTE modeling of 
the \ion{Li}{1} feature yielded a redshifted line with respect to 1D LTE results, inconsistent
with the shift we observed in MML 28.  Given the doublet structure of the ${\lambda}6708$ feature,
it is possible that the magnetic intensification mechanism described by \citet{Leone2007} could
strengthen the blue component of the resonance line relative to the red component.  However, our
results indicate that MML 28 does not evince any line strength enhancement, relative to other
stars, that should accompany such a wavelength shift relative to other stars in the sample. 
There is, however, observational evidence \citep{AP98,R02,Mandell04} that convective motions can 
produce \ion{Li}{1} resonance lines with blueshifted centroids.  In the 6708 {\AA} synthesis for 
MML 28, we have simply shifted the synthetic Li lines by -0.016 {\AA}, roughly 2-3 times the shift 
of the stronger Li hyperfine components applied by \citet{Mandell04} in their linelist calibrated 
with the Sun.  

Input abundances for the initial syntheses were solar values scaled according to the [Fe/H] 
for each star derived by \citet{epaper}.  Small adjustments (a couple hundredths of a dex) 
within the uncertainties of \citet{epaper} were allowed if they consistently improved the 
fitting of blending features neighboring both the 6104 and 6708 {\AA} Li features.  Smoothing 
was accomplished utilizing a Gaussian with a fixed (in each wavelength region) FWHM corresponding 
to the instrumental resolution ($R{\sim}40,000$) and then employing rotational broadening which consistently 
yielded the best fit to the line profiles in both the 6104 and 6708 {\AA} region.  As expected, the \vsini\, values 
we determined in this fashion are (except for MML 30 and MML 73) slightly smaller than the values measured 
by \citet{epaper} that parameterized the total broadening (instrumental, thermal, rotational, etc). 

Representative observed spectra and syntheses (including our best fit 
values) are shown in Figures 1 and 2 for the 6104 and 6708 {\AA} regions respectively. 
Comparison of our spectra with syntheses having no Li clearly and confidently indicates the 
presence of the blended, moderate strength ${\lambda}6104$ \ion{Li}{1} feature in all but one 
of our stars' spectrum (MML 73).  For this star, our sense is that the ${\lambda}6104$ is present and measurable, 
but 2 uncooperative pixels in the midst of the blended Li feature do not make this conclusion definitive 
(see Figure 3).  Our ${\lambda}6104$ abundance is included in all of the statistical
analyses (unless stated otherwise), but the star is denoted as an inverted triangle in the figures. The best 
fit LTE Li abundances are provided in columns 2 and 4 of Table 2.  
\marginpar{Fig1}
\marginpar{Fig2}
\marginpar{Fig3}
\marginpar{Tab2}

NLTE corrections were applied to both the ${\lambda}6104$ and ${\lambda}6708$ values according to the 
prescription of \citet{Carlsson1994}.  Interpolations within their grid of NLTE corrections provided 
the necessary correction to our LTE lithium abundances, based on 
$T_{\rm eff}$, log $g$, [Fe/H], and log $N$(Li).  Three stars had absolute physical parameters which 
placed them outside the range of values populated by the grid.  Within the adopted uncertainties of 
the parameters, however, the stars all fall within an acceptable extrapolation range.  For MML 72, 
which had a supersolar metallicity of 0.06, we adopted a correction which
assumed solar metallicity, justified based on the apparent insensitivity of the grid to changes in
[Fe/H] (e.g. changes in metallicity of -0.71 dex and -0.14 dex are required to alter the 6104 and 6708 {\AA}
NLTE corrections by 0.01 dex at the $T_{\rm eff}$, logg and Li abundances characterizing the star).  
Both MML 43 and MML 44 had surface gravities which were outside of the range covered by
the \citet{Carlsson1994} grid.  For these stars we assumed a surface gravity of 4.5, the maximum value in the grids.  
The sensitivity of the NLTE corrections to changes in the surface gravity is small:{\ \ }a -0.20 dex difference 
in gravity for both MML 43 and MML 44 results in a ${\le}$+0.01 dex difference in the NLTE corrections for the 
6708 {\AA} line, and log $g$ changes of ${\ge}1.0$ dex are needed to observe a 0.01 dex change in NLTE 
corrections for the 6104 {\AA} line).  This suggests that surface gravity 
changes have a negligible effect on the NLTE corrections, given the uncertainties.  
Final NLTE Li abundances are provided in columns 
3 and 5 of Table 2.  

Uncertainties in the Li abundances are dominated by two general components:{\ \ }uncertainties
in the fitting and the $T_{\rm eff}$ values (uncertainties in log $g$ and ${\xi}$ contribute negligibly). 
The fitting uncertainties are dominated by uncertainties in the continuum location and \vsini\, 
values.  Plausible necessarily subjective estimates of allowed variations in these quantities 
were made and the Li abundances rederived to estimate the fitting uncertainty.  These fitting 
uncertainties were added in quadrature with the abundance uncertainties due to the uncertainty in 
$T_{\rm eff}$ from \citet{epaper} to estimate the total uncertainty in our Li abundance from each 
of the 6104 and 6708 {\AA} features.  These are listed in columns 3 and 5 of Table 2.  We note
that the uncertainties in the ${\lambda}6104$- and ${\lambda}6708$-based Li abundances due to 
T$_{\rm eff}$ uncertainty are correlated.  When we later examine the difference between the Li
abundances derived from the two wavelength regions, we account for the correlation of this component of 
the uncertainties in calculating the uncertainty of the difference.

\section{RESULTS AND DISCUSSION}

\subsection{Lithium Dispersion}

The 6104 {\AA}- and 6708 {\AA}-based NLTE Li abundances are plotted versus $T_{\rm eff}$ in 
Figure 4.  In traditional open cluster work, the $T_{\rm eff}$ coordinate serves as a monotonic
mass coordinate, but this does not hold in very young star formation regions/associations such
as Sco-Cen.  Accordingly, we plot the NLTE Li abundances versus mass in Figure 5.  We utilized 
bolometric luminosities from \citet{Mamajek2002}
and spectroscopic effective temperatures to interpolate within the pre-main sequence grids of 
\citet{DM97} in order to derive masses for our stars.  It should be noted that our track-based masses
may be underpredicted.  According to \citet{Hillenbrand2004}, masses from
evolutionary tracks underpredict PMS stellar masses by as much as 10-30 \% when compared with dynamical
masses.  Such an offset could presumably account for some of the lack of observed depletion, but
quantifying this is difficult with currently available models.  Consequently we proceed with
the analysis making use of our derived masses.
\marginpar{Fig4}
\marginpar{Fig5}

The ordinary Pearson correlation coefficients indicate no significant variation in either
Li abundance with stellar mass.  A striking result is the pleasingly small scatter in the 
${\lambda}6104$ \ion{Li}{1} abundances compared to the ${\lambda}6708$-based values.  The
smaller scatter in the former is consistent with the smaller uncertainties we find for the
former.  Historically, the ${\lambda}6104$ line has not frequently been utilized in abundance
studies due to its weakness and increased blending compared to the ${\lambda}6708$ resonance
feature.  Our results, however, suggest that the subordinate feature may produce higher 
quality abundances than the resonance feature in young Li-rich stars.    In addition, the NLTE 
corrections are significantly less sensitive to changes in both metallicity and surface gravity for 
the 6104 line.  Independent changes in log $g$ of $\sim$$\pm$0.20 dex and $\sim$$\pm$0.15 dex in [Fe/H] 
yield changes of $\sim$$\pm$0.01 dex in the NLTE
corrections for the 6708 line.  A change of order 1.0 dex and 0.71 in log $g$ and [Fe/H], respectively,
is needed to see a 0.01 dex change in the NLTE correction for the 6104 line.  
Based on the lower scatter from the 6104 {\AA} line and the lower sensitivity to parameter changes 
(log $g$ and [Fe/H]) for NLTE corrections to the subordinate line, we conclude, similarly to \citet{Carlsson1994} 
(see Figure 10 of that work), that the 6104 {\AA} feature is more well-suited for obtaining lithium abundances 
in young, lithium rich solar-type stars.

The rms scatter in the ${\lambda}6104$- ${\lambda}6708$-based Li abundances about their
respective means in Figure 5 (0.15 and 0.25 dex, respectively) compares favorably with the typical 
uncertainties (0.08 and 0.18 dex, respectively).  The ${\lambda}$6708-based scatter is dominated 
by the lower abundance of MML 73; though nothing else appears remarkable or notably different 
about this star, the ${\lambda}$6708-based rms scatter of 0.17 dex found when excluding it is 
equivalent to that expected on the basis of the uncertainties alone.  We conclude 
from these comparisons that any real Li dispersion--measured by the square root of the difference 
between the variance of the Li abundance data and the square of typical Li abundance uncertainty--in 
our sample of modestly rotating stars is limited to ${\le}0.15$ dex.  

The ${\lambda}6708$- and ${\lambda}6104$-based Li abundance of cool slowly-rotating Pleiads with 
$T_{\rm eff}{\le}5500$ K,  which characterizes most of the stars in our Sco-Cen sample,  show a real 
dispersion of ${\sim}0.6$ dex \citep{KSHP10}.   Comparable spreads  in ${\lambda}6708$-based 
abundances are seen for $T_{\rm eff}{\le}5200$ K in the 50-90 Myr  ${\alpha}$ Per cluster
\citep{Bala96,Stauffer1999,Bala2011}.  
However, mass is the relevant variable in considering PMS Li depletion, and the masses of the Pleiads 
that evince significant dispersion are some 0.3-0.6 $M_{\odot}$ lower in mass than the stars in our 
sample.  Pleiads having masses comparable to those objects in our sample show no significant 
Li dispersion \citep[e.g., Figure 4 of][]{K2000}.   Our data therefore do not address the interesting question 
of whether the Pleiades and ${\alpha}$ Per Li dispersion is the relic of possible differential Li depletion 
mechanisms having acted at the ${\sim}15$ Myr age of Sco-Cen.  

\subsection{$\lambda$6104 versus $\lambda$6708}

Figure 6 shows the difference between the NLTE ${\lambda}6708$- and ${\lambda}6104$-based Li abundances
versus $T_{\rm eff}$.  The mean difference (6708 minus 6104) is $+0.20{\pm}0.07$ (mean uncertainty) dex.  
The 7 stars with $T_{\rm eff}{\ge}5200$ K show no significant difference in the Li abundances derived
from the two Li features, a result consistent with the abundances derived in similarly warmer Pleiads 
\citep{KSHP10}.  Figure 6 indicates that the stars with statistically significant 0.3-0.5 dex abundance 
differences are the coolest objects in our sample. 
\marginpar{Fig6}

The behavior of this difference may qualitatively fit the picture of magnetic intensification suggested
by \citet{Leone2007}, where magnetic fields of 10$^{-1}$ to 1 T can enhance the strength of 
strong \ion{Li}{1} resonance features relative to weak ones.  Whether such an effect would:{\ }a) hold 
for the ${\lambda}6104$ feature, (b) hold for stars of our $T_{\rm eff}$, gravity, and Li abundance, and
(c) would ``switch on'' at $T_{\rm eff}{\le}5200$ K as we observe, all require additional calculations 
beyond those carried out by \citet{Leone2007}.  

Recent works point to the existence of abundance patterns consistent with overexcitation/ionization in cool 
($T_{\rm eff}{\le}5200$ K) open cluster dwarfs \citep{Sch03,Sch04,Sch06}.  However, such a mechanism 
(whatever its cause[s]) acting in the framework of LTE can not explain the difference and the sense
of the difference we observe in our coolest stars.  Figure 2 provides an important clue to the source
of the behavior in demonstrating that the ${\lambda}6708$-${\lambda}6104$ difference is due 
to declining ${\lambda}6104$ abundances (and not increasing ${\lambda}6708$ abundances) with 
declining $T_{\rm eff}$.  We believe that this is a signature of relative NLTE overphotoionization 
at low $T_{\rm eff}$ acting to selectively depopulate the 2\emph{p} level (the lower level of the ${\lambda}6104$
feature relative to the 2\emph{s} level (the lower level of the ${\lambda}6708$ feature), thus
weakening the ${\lambda}6104$ line relative to the ${\lambda}6708$ line.  Indeed, \citet{Carlsson1994} 
note the importance of ultraviolet photoionization as a NLTE mechanism important for \ion{Li}{1}
line formation and PMS stars exhibit greater UV emission \citep{Findeisen2010}.  
A significant finding of their work is that, at low $T_{\rm eff}$ and high Li 
abundance, (near)ultraviolet metal line opacity inhibits photoionization from the Li 2\emph{s} level, 
allowing photoionization from the 2p level to dominate.  

Our results and those of \citet{KSHP10} indicate that the relative ${\lambda}6104$ versus 
${\lambda}6708$ \ion{Li}{1} NLTE corrections from \citet{Carlsson1994} are adequate for young Li rich stars 
with $T_{\rm eff}{\ge}5200$ K, but the ${\lambda}6104$ corrections are too small, perhaps by 
${\ge}0.4$ dex, at cooler $T_{\rm eff}$. This suggests that while the internal uncertainties and sensitivity 
of NLTE corrections to stellar parameters is smaller for the ${\lambda}6104$ feature relative to
the ${\lambda}6708$ feature, the absolute abundances from the ${\lambda}6708$ may be more reliable 
modulo other possible (perhaps activity-related) effects on the resonance line \citep{KS04}. 

\subsection{Sco-Cen Lithium Abundance}

Since the ${\lambda}6104$ \ion{Li}{1} abundances are likely not free of systematic errors in the 
NLTE corrections for the cooler stars as discussed above, our mean Li abundance estimate of 
log $N$(Li)$=3.50{\pm}0.08$ (mean uncertainty) is made from the ${\lambda}6708$-based NLTE results. 
This abundance is ${\sim}2{\sigma}$ larger than the meteoritic value.  While no mass-dependent 
slope to the Li data indicative of depletion is seen in Figure 5, even the most conservative
PMS Li depletion models predict depletion factors of ${\ge}0.4$ dex for 1 $M_{\odot}$ stars at 
an age of 15 Myr \citep{PT02} (see their Figure 4).  Our results suggest the initial Li abundance of 
our cooler objects might be a factor of 2-3 higher than observed today.  On the other hand, the
existence of PMS Li depletion inhibiting mechanisms remains possible \citep{VZM98}.  Assessing 
whether the degree of Li depletion suffered by our Sco-Cen sample and whether our data are in 
agreement with PMS Li depletion models will require extending the Li-mass relation via observation 
and analyses of cooler, lower mass Sco-Cen members that are expected to be more prodigous depleters.  

\acknowledgments
EJB and EEM acknowledge support from NSF grant AST 10-08908 to EEM and the University of Rochester School of Arts and Sciences.
MS and JRK gratefully acknowledge support for this work from NSF grant AST 09-08342 to JRK.  We would also like to acknowledge
the referee whose comments helped to clarify the work.

\begin{figure}
\includegraphics[width=3.5in]{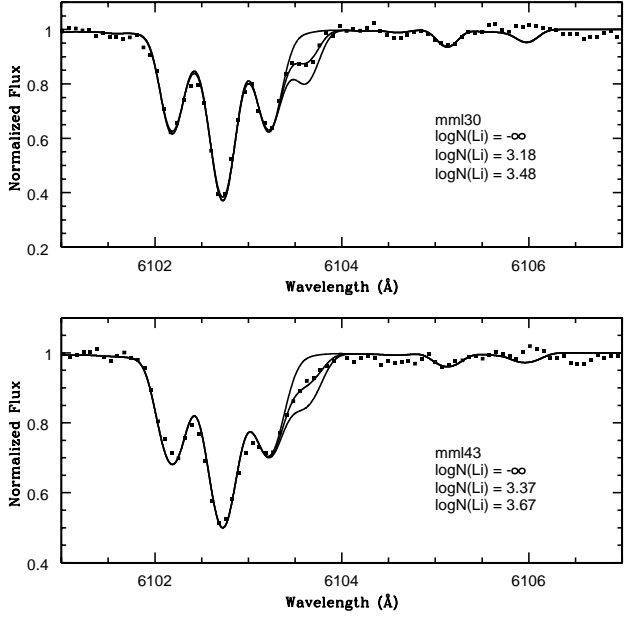}
\caption{Sample lithium syntheses of the 6104 {\AA} subordinate line of
lithium for MML 30 (top) and MML 43 (bottom).  The observed spectrum is plotted as
solid squares and three syntheses are shown for each star.}
\label{figure:mml30_43_6104}
\end{figure}

\begin{figure}
\includegraphics[width=3.5in]{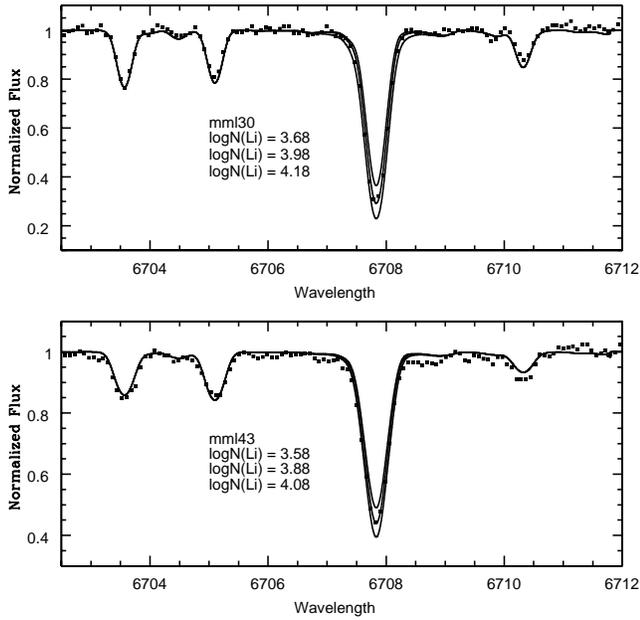}
\caption{Sample lithium syntheses of the 6708 {\AA} doublet of Li for
MML 30 (top) and MML 43 (bottom).  The observed spectrum is plotted as
solid squares and three lithium syntheses are shown for each star.}
\label{figure:6708}
\end{figure}

\begin{figure}
\includegraphics[width=3.5in]{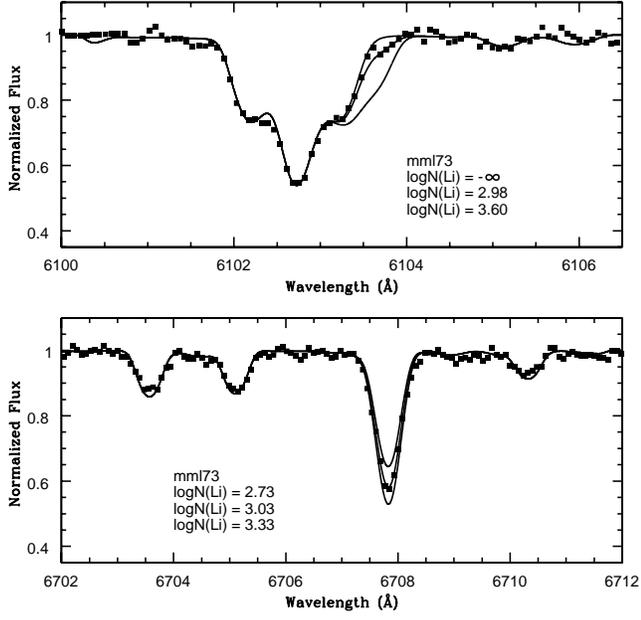}
\caption{Sample lithium syntheses of the $\lambda$6104 line of lithium (top)
and the $\lambda$6708 doublet (bottom) for MML 73.  This measurable abundance 
of lithium from the 6104 {\AA} feature is less definitive in this star than in the rest of 
our sample.  The 6708 {\AA} results, however, show the clear presence of Li in the
star.} 
\label{figure:mml73}
\end{figure}

\begin{figure}
\includegraphics[width=3.5in]{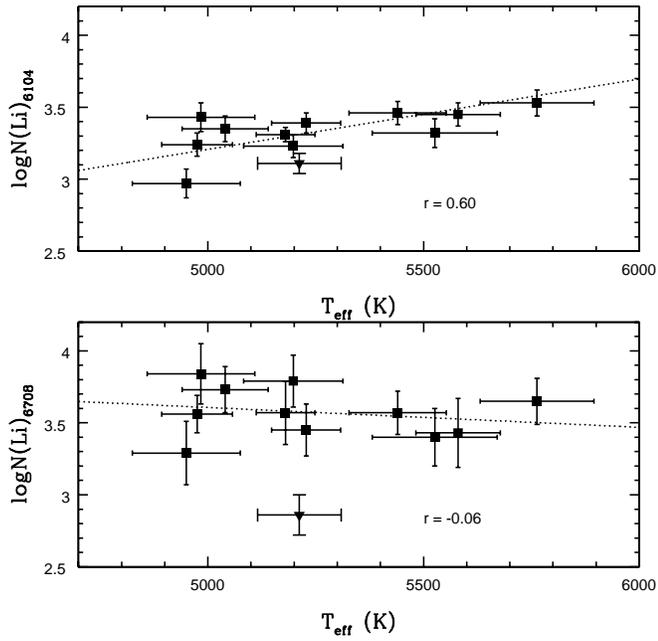}
\caption{The NLTE corrected abundance of lithium is plotted versus 
temperature for the stars based on the $\lambda$6104 subordinate
feature (top) and the $\lambda$6708 resonance doublet (bottom).
The least squares fit performed in each case is shown as the dotted 
line, and the ordinary Pearson correlation coefficient between Li abundance
in $T_{\rm eff}$ is labeled in the plots.}
\label{figure:scatter}
\end{figure}

\begin{figure}
\includegraphics[width=3.5in]{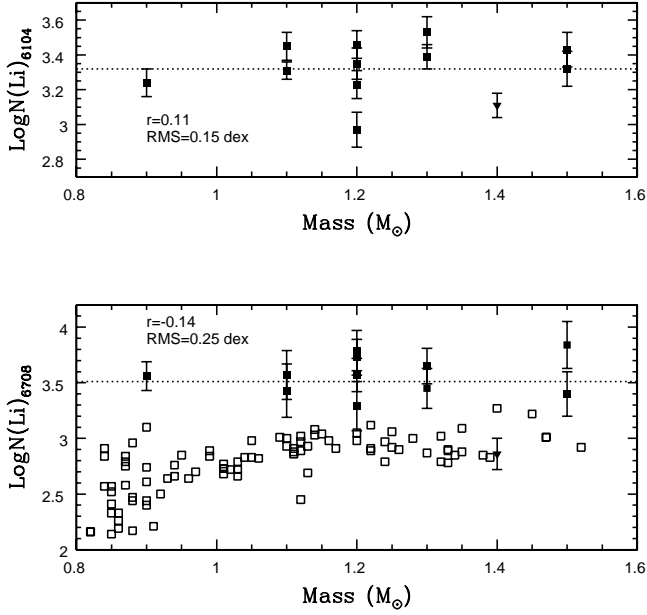}
\caption{The NLTE abundances, logN(Li)$_{6104}$ (top)
and logN(Li)$_{6708}$ (bottom), are plotted versus mass.  NLTE corrected
abundances for a sample of Pleiades stars
\citep[from ][]{Soderblom1993} are plotted for logN(Li)$_{6708}$ 
(bottom-open squares) as a comparison sample which exhibits depletion.  Note that
our lowest mass star clearly shows no evidence of depletion, which is contrary to
stellar lithium depletion models given the age and assumed mass.}
\label{figure:nlte}
\end{figure}

\begin{figure}
\includegraphics[width=3.5in]{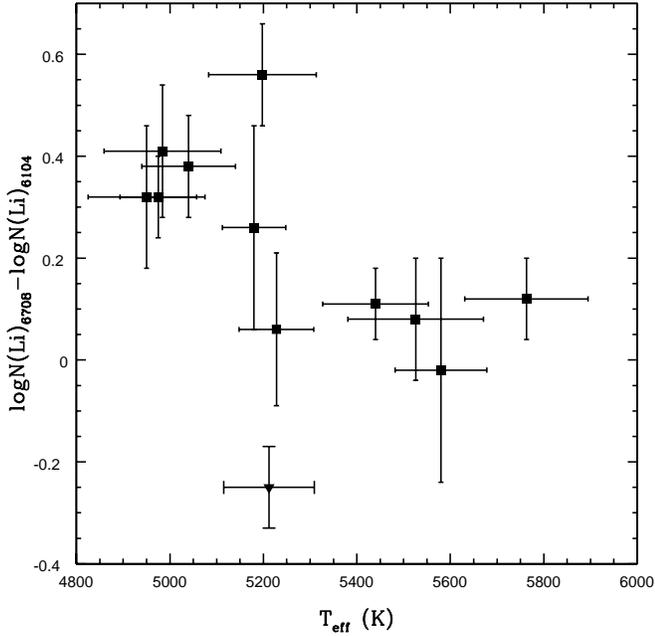}
\caption{The difference in NLTE abundances, logN(Li)$_{6708}$-logN(Li)$_{6104}$ is 
plotted versus temperature.  The difference that is seen at T$_{eff}\le$5200 K
is due to the effects of overphotoionization on $\lambda$ 6104 which manifests itself
by yielding lower abundances when compared with the $\lambda$ 6708-based abundances.}
\label{figure:diff}
\end{figure}

\begin{deluxetable}{l c c c c c c}
\tablewidth{0 pt}
\tablecaption{Physical Parameters}
\tablehead{
\multicolumn{1}{l}{Name} &
\multicolumn{1}{c}{Teff} &
\multicolumn{1}{c}{logg} &
\multicolumn{1}{c}{[Fe/H]} &
\multicolumn{1}{c}{MtVel} &
\multicolumn{1}{c}{log($\frac{L}{L_{\odot}}$)} &
\multicolumn{1}{c}{Mass} \\
\multicolumn{1}{l}{} &
\multicolumn{1}{c}{K} &
\multicolumn{1}{c}{} &
\multicolumn{1}{c}{} &
\multicolumn{1}{c}{\kms} &
\multicolumn{1}{c}{dex} &
\multicolumn{1}{c}{M$_{\odot}$}\\
}
\startdata
MML 7  & 5763 $\pm$   132   &  4.24 $\pm$ 0.34 & -0.01 $\pm$ 0.11 & 2.54 $\pm$ 0.14 & 0.39 $\pm$ 0.06 & 1.3 \\
MML 13 & 4984 $\pm$   125   &  3.99 $\pm$ 0.54 & -0.27 $\pm$ 0.10 & 2.53 $\pm$ 0.28 & 0.26 $\pm$ 0.10 & 1.5 \\
MML 28 & 4975 $\pm$   82    &  4.08 $\pm$ 0.40 & -0.21 $\pm$ 0.05 & 2.08 $\pm$ 0.16 & -0.37 $\pm$ 0.14 & 0.9 \\
MML 30 & 5180 $\pm$   68    &  4.02 $\pm$ 0.30 & -0.13 $\pm$ 0.05 & 2.18 $\pm$ 0.12 & -0.05 $\pm$ 0.09 & 1.1 \\
MML 36 & 5198 $\pm$   115   &  4.39 $\pm$ 0.29 & -0.09 $\pm$ 0.09 & 1.96 $\pm$ 0.22 &  0.00 $\pm$ 0.05 & 1.2 \\
MML 40 & 5440 $\pm$   113   &  4.36 $\pm$ 0.20 & -0.22 $\pm$ 0.09 & 3.16 $\pm$ 0.22 &  0.07 $\pm$ 0.08 & 1.2 \\
MML 43 & 5580 $\pm$   98    &  4.60 $\pm$ 0.28 & -0.07 $\pm$ 0.08 & 2.59 $\pm$ 0.25 &  0.06 $\pm$ 0.08 & 1.1 \\
MML 44 & 5526 $\pm$   145   &  4.71 $\pm$ 0.36 & -0.05 $\pm$ 0.12 & 2.29 $\pm$ 0.29 &  0.54 $\pm$ 0.10 & 1.5 \\
MML 55 & 5228 $\pm$   80    &  4.28 $\pm$ 0.24 & -0.12 $\pm$ 0.07 & 2.50 $\pm$ 0.16 &  0.16 $\pm$ 0.11 & 1.3 \\
MML 70 & 5040 $\pm$   100   &  4.09 $\pm$ 0.20 & -0.18 $\pm$ 0.10 & 2.24 $\pm$ 0.15 &  0.01 $\pm$ 0.10 & 1.2 \\
MML 72 & 4950 $\pm$   125   &  4.19 $\pm$ 0.44 &  0.06 $\pm$ 0.13 & 1.31 $\pm$ 0.42 &  0.21 $\pm$ 0.07 & 1.2 \\
MML 73 & 5212 $\pm$   97    &  4.00 $\pm$ 0.32 & -0.20 $\pm$ 0.09 & 2.42 $\pm$ 0.19 &  0.21 $\pm$ 0.01 & 1.4 \\
\enddata
\label{phys_param}
\end{deluxetable}

\begin{deluxetable}{l c c c c}
\tablewidth{0 pt}
\tablecaption{Lithium Abundance}
\tablehead{
\multicolumn{1}{l}{Name} &
\multicolumn{1}{c}{logN(Li)$_{6104}$} &
\multicolumn{1}{c}{logN(Li)$_{6104}$} &
\multicolumn{1}{c}{logN(Li)$_{6708}$} &
\multicolumn{1}{c}{logN(Li)$_{6708}$} \\
\multicolumn{1}{l}{} &
\multicolumn{1}{c}{LTE} &
\multicolumn{1}{c}{NLTE} &
\multicolumn{1}{c}{LTE} &
\multicolumn{1}{c}{NLTE} \\
}
\startdata
MML 7    & 3.47        & 3.53 $\pm$ 0.09 & 4.22  & 3.65 $\pm$ 0.16 \\
MML 13   & 3.28        & 3.43 $\pm$ 0.10 & 4.15  & 3.84 $\pm$ 0.21 \\
MML 28   & 3.08        & 3.24 $\pm$ 0.08 & 3.93  & 3.56 $\pm$ 0.13 \\
MML 30   & 3.18        & 3.31 $\pm$ 0.05 & 3.98  & 3.57 $\pm$ 0.22 \\
MML 36   & 3.11        & 3.23 $\pm$ 0.08 & 4.14  & 3.79 $\pm$ 0.18 \\
MML 40   & 3.37        & 3.46 $\pm$ 0.08 & 4.04  & 3.57 $\pm$ 0.15 \\
MML 43   & 3.37        & 3.45 $\pm$ 0.08 & 3.88  & 3.43 $\pm$ 0.24 \\
MML 44   & 3.24        & 3.32 $\pm$ 0.10 & 3.84  & 3.40 $\pm$ 0.20 \\
MML 55   & 3.27        & 3.39 $\pm$ 0.07 & 3.83  & 3.45 $\pm$ 0.18 \\
MML 70   & 3.20        & 3.35 $\pm$ 0.09 & 4.07  & 3.73 $\pm$ 0.16 \\
MML 72   & 2.81        & 2.97 $\pm$ 0.10 & 3.59  & 3.29 $\pm$ 0.22 \\
MML 73   & $\le$2.98   & 3.11 $\pm$ 0.07 & 3.03  & 2.86 $\pm$ 0.14 \\
\enddata
\label{liabundance}
\end{deluxetable}

\end{document}